\documentstyle[aps]{revtex}

\begin{document}

\title{Weak Quantum Ergodicity}

\author { L. Kaplan
\thanks{kaplan@phyics.harvard.edu}}
\address{Department of Physics and Society of Fellows,\\ Harvard
University\\ Cambridge, Massachusetts 02138\\}

\author { E. J. Heller\thanks{heller@physics.harvard.edu}}
\address{Department of Physics, Harvard University
and\\ Harvard-Smithsonian Center for Astrophysics\\ Cambridge,
Massachusetts 02138\\}
\date{\today}
\maketitle

\begin{abstract}
We examine the consequences of classical ergodicity for the localization
properties of individual quantum eigenstates in the classical limit.
We note that the well known Schnirelman result is a  weaker
form of quantum ergodicity than the one implied by random matrix theory.
This suggests the possibility of systems with non-gaussian random eigenstates
which are nonetheless ergodic  in the sense of Schnirelman and lead to
ergodic transport
in the classical limit. These we call ``weakly quantum ergodic.''
Indeed for a class of ``slow ergodic" classical systems, it is found that  each eigenstate
becomes localized to an ever decreasing fraction of the available state
space, in the semiclassical limit.
Nevertheless, each eigenstate in this limit covers phase space evenly on
any classical scale, and long-time
transport properties betwen individual quantum states remain ergodic due to the
diffractive effects which dominate
quantum phase space exploration.
\end {abstract}

\section{Introduction}

The issue of quantum correspondence to classical ergodicity
has received much attention in recent years.
Important questions surround the properties  of individual quantum eigenstates
of  classically ergodic systems.
How do eigenstates reflect the ergodic nature of the
underlying classical dynamics,
in the classical limit?  Must they be individually ergodic?
What latitude is there in the eigenstates in order that
they support classical ergodicity in the correspondence limit?
There is a history of interesting conjectures,
numerical results, theories, and even theorems in this field.  Our
focus here is on ergodic systems which are not chaotic in the
sense of positive Lyapunov exponents, as these systems spawn the
unusual eigenstate properties uncovered in this paper.

A natural place to begin is with two important  conjectures.
In 1983  Berry suggested that eigenstates of classically ergodic
systems locally look like random superpositions of plane waves\cite{berry1}.
Essentially,
``random superposition''  here means the
addition of plane waves with random direction, amplitude,  and phase,
all having the same local wavelength.  The central limit theorem
makes it unimportant what distribution the amplitudes are drawn
from,
as long as they are truly random and independent.

The second landmark conjecture is due to Bohigas, Giannoni, and
Schmit\cite{BGS}.  They
suggested that spectra (and eigenstate properties) of classically chaotic
Hamiltonian systems would be identical to those of random matrix theory
(RMT)\cite{rmt}
(see also Ref.\cite{hellerRMT}) in the
classical limit.  This
implies strong level repulsion in the spectrum
and Gaussian random eigenfunctions, with
detailed characteristics
of both varying according to symmetry properties of the system
(e.g. the presence or absense of time
reversal
symmetry).

Subsequent numerical and theoretical work has borne out these conjectures
in large part,
with modifications that can   be ascribed to short time classical dynamics.
Classical chaos in the presence of a typical Lyapunov exponent
$\overline\lambda>0$ takes on the order of the ``log time,''
$t = |\log \hbar|/{\overline\lambda}$,
to spread a
Planck-sized cell everywhere allowable in phase space on a Planck-scale
mesh (at which time the cell
has lost all memory of its initial location)\cite{log}.
This is associated with
the positive entropy of the chaotic dynamics, which destroys the
information contained in the initial wavepacket. Following the log-time
the classical dynamics begins to create structures in phase-spcae on
sub-Planck scales, which get washed out quantum mechanically by
the uncertainty principle. Before this time,
the imprint of nonrandom classical mechanics is written or encoded into the
quantum mechanics, so to speak.
Time dependent semiclassical approximations
hold for a time at least on the order of the log time\cite{log}.  Since the
quantum eigenstates (and their eigenvalues)  support  the short-time
as well as the long-time dynamics, they
must possess
corrections  to RMT.
 However, since it takes a much longer time
(the Heisenberg time, or $\hbar/\delta E$, where $\delta E$ is the
mean level spacing)
to resolve individual eigenstates, the corrections to universal RMT behavior
often involve collective properties of many nearby eigenvalues and their
corresponding eigenstates.

These corrections
stand out against the monotonous  backdrop of RMT.
For example, some eigenstates have enhanced amplitude near,
i.e. are ``scarred'' by, certain short isolated unstable periodic orbits.
 Although random eigenstates (in the sense of
Berry's conjecture) have scar-like concentrations when viewed in coordinate
space\cite{gehlen},
eigenstates of chaotic systems have scars   associated with definite
periodic orbits \cite{scar}.
 The   theory of scars\cite{scar,Bogoscar,berryscar}
 depends only on short time, linearized dynamics near the periodic
orbit.
These  are ``linear'' scar theories,
they do not tell us {\it which } states are scarred or what individual
states look like.
Nonlinear scar theories provide information about how
scar density from a given orbit is distributed among individual
eigenstates\cite{steve}; these theories
are still under development\cite{tocome}.

Other corrections to  RMT
 involve nonuniversal correlations in eigenvalue
spectra, which also arise from  short  periodic orbits\cite{berry}.
 Much numerical
evidence supports the existence of scars and nonuniversal spectral
correlations.

Given that deviations from RMT are known to exist, the question arises
as to how strong a form of quantum ergodicity is actually required for
a system which is classically ergodic. Naively, classical--quantum
correspondence only has implications for  properties averaged over many eigenstates,
such as transport efficiency from one region of phase space to another.  What are the 
constraints imposed  on individual eigenstates?  

A very important result in this regard was proved
by Schnirelman\cite{schnirl} and
Colin de Verdiere\cite{colin}
 (see also Zelditch\cite{zel}).  The result
states that in the classical limit, the quantum expectation value of certain
operators
\footnote{Specifically, Colin de Verdiere's proof
deals with pseudo-differential operators,
acting on eigenstates of the Laplacian
on a compact space without boundaries. Zelditch's work extends these results
to billiards with piecewise smooth walls, with the condition that the kernel
of the operator must be supported away from the walls. Others have dealt
with the case of a compact classical phase space\cite{cptspace}.}
over {\it individual} eigenstates
is almost always the ergodic, microcanonical average of the classical
version of the operator,
 proving the ergodicity of individual eigenstates on a ``macroscopic'' scale.
Colin de Verdiere's proof left a loophole for a possible  measure zero set
of states
which could violate   ergodicity; he conjectured however
that the exceptions did not exist.
It was  shown later that a strongly phase space localized
(but zero measure as $\hbar\to 0$) set
of states could indeed exist\cite{bounce}.
In effect, Schnirelman-Colin de Verdiere-Zelditch (SCdVZ) showed
that eigenstates of classically ergodic systems
cannot deviate from ergodicity when smoothed over
any finite patch of phase space as  $\hbar\to 0$.
Importantly,
the presence of macroscopic, classical scale ergodicity leaves open the
question of microscopic ergodicity, on classically infinitesimal (but
quantum mechanically large) scales.

  An example of
a possible ``microscopic''  scale is a length scaling as $\sqrt{\hbar}$.
This length, although becoming (as  $\hbar\to 0$) insignificant on any scale
over which a smooth classically-defined
operator can vary, contains
infinitely many de Broglie wavelengths in the same limit.  One might suppose
that eigenstates of classically chaotic systems are generally ergodic
even on these microscopic scales, but this remains to be shown.  Eigenstates
behaving in accordance with RMT would be microscopically
ergodic, forbidding large deviations in $\vert\psi\vert^2$ when averaged over
a $\sqrt{\hbar}$ length scale (or indeed over any
scale large compared to $\hbar$)
as $\hbar\to 0$.
To put it another way, {\it random matrix theory is a far more stringent
requirement than}  SCdVZ  {\it ergodicity.}
It seems appropriate to call SCdVZ ``weak quantum ergodicity'' :
eigenfunctions which are not Gaussian random in most representations can still
be ergodic
in the sense of  SCdVZ. A weakly quantum ergodic  eigenstate can
become localized to an ever decreasing fraction of the available state space,
in the classical limit,
but in such a way that the coarse graining provided by taking the expectation
value of any
classically
smooth operator leads to a result which is not anomalous.
This paper shows that the regime of weak quantum ergodicity
exists, and provides examples.

We  first develop a general formalism for analyzing and comparing various
types of
localization phenomena. Then   the implications of classical
ergodicity for the properties of quantum eigenstates and quantum transport
are discussed, and compared with random
matrix theory (RMT) predictions. Several known examples of anomalous
(non-RMT) behavior in
classically ergodic systems  are mentioned. It is shown how such anomalies
(and indeed even more striking ones) are
consistent with the predictions of Schnirelman-Colin de Verdiere.
We examine the
important effect that
classical spreading rates have on quantum behavior,
and on the relative importance of classical and diffractive
phase space exploration. Finally, we discuss two examples of weakly quantum
ergodic systems.

\section{Localized Eigenstates and Ergodic Quantum Transport}

The question of quantum ergodicity for conservative
Hamiltonian systems suffers from the following conceptual problem:
energy is conserved in both classical and quantum mechanics, and is a known
{\it a priori} constraint on the dynamics.  In classical mechanics this is
easily handled
by considering ergodicity on the energy hypersurface; in quantum mechanics
it would beg the question to do so, since the only states with fixed energy
are eigenstates.  These are unique unless the spectrum is degenerate. Thus, if
we insist on working on an energy hypersurface, all  nondegenerate systems
are trivially
ergodic.  This  in fact was the basis of von Neumann's approach to
quantum ergodicity.  It has
the problem of an absence of classical-quantum correspondence
for nondegenerate systems.
The problem is resolved by considering flow between phase space localized
states
(usually coherent states)\cite{ergodic}. These of course populate a range
of energies, but
the moments of the energy distribution for any such state
are known from very short time
dynamics.  Then
it is straightforward to develop criteria for ergodic flow between  such
localized states\cite{ergodic}. 

If energy is not conserved (as in a kicked system) the problem of
incorporating the
{\it a priori} constant energy constraints does not arise\footnote{
Of course any non-autonomous system is equivalent to an autonomous one
in an extended phase space. In this extended phase space, the ergodic
dynamics would only produce uniform coverage on the constant-``energy"
hypersurface, as discussed in the preceding paragraph.}.
The situation is also conceptually
simpler  in   any space where for some reason
each basis state is {\it a priori} equally accessible.  An example
is a kicked symmetric top where the kick preserves total angular
momentum $J$  and body-axis angular momentum $K$
(and thus the energy) but not  $M$, the space fixed $z-$component of
angular momentum. Then all $(2J + 1)$ $M$ states are degenerate and are equal
candidates for
flow under the
dynamics. The semiclassical limit is $J \to \infty.$
Every  $M$ state in the space can be made equally  accessible
 starting from any other state, or any superposition of states.  We will
formulate the
transport theory and our models along these lines, but   weakly quantum
ergodic systems
will exist more generally, even in the presence of constraints imposed
by classical symmetries.

The problem of localization {\it vs.} ergodicity may be stated as
the problem of the distribution of
overlap probabilities
\begin{equation}
p_{n}^a = |\langle a|n\rangle |^2\equiv |\alpha_n^a|^2,
\end{equation}
where $|a\rangle $ is some  physically appropriate basis
spanning an eligible  subspace,
 and $|n\rangle $  are eigenstates of the propagator (e.g. a kick)
spanning the same space. ``Eligible'' means states that are accessible to the
dynamics according to
all {\it a priori} constraints. The probabilties are normalized,
i.e.
\begin{equation}
\sum\limits_n p_{n}^a = \sum\limits_a p_{n}^a = 1.
\end{equation}
If the propagator is a discrete quantum unitary map $S$, the eigenstates
$\vert n\rangle$ are
specified  by
\begin{equation}
S \vert n\rangle = e^{i \phi_n} \vert n\rangle \,,
\end{equation}
where the real phase $\phi_n$ puts all the eigenvalues on the unit circle.
If the total Hamiltonian can
be   written as $H = H_0 + \delta H$, it may be interesting to use the
eigenstates of $H_0$ as the
basis $|a\rangle $, and to consider how these are intermixed by the
perturbation $\delta H$. For example, for a
particle hopping on a random lattice, as in the case of Anderson localization,
the position basis may be a natural
one to use. There we can imagine starting with random on-site potentials
and adding nearest-neighbor couplings as a perturbation. In the scattering and
kicked systems  discussed towards the end of this paper, we will take the
reference basis $|a\rangle $ to be
  momentum eigenstates (or channels in the case of scattering).
 Other bases, such as phase space bases, may also be important in
various contexts (e.g. in the presence of constraints imposed by time
translation invariance or other symmetries).

Under   discrete time dynamics,
governed by  a unitary operator $S$, a state $\vert a \rangle$ becomes
after $\ell$ iterations
\begin{equation}
\vert a_\ell \rangle = S^\ell \vert a \rangle = \sum\limits_n \exp [ i \ell
\phi_n] \alpha_n^a \vert n \rangle \,,
\end{equation}
so that   in the nondegenerate
case the time averaged density corresponding to the initial state $\vert
a\rangle$ becomes
\begin{eqnarray}
\rho_a^\infty &\equiv& \lim\limits_{L\to\infty}{1\over L}
\sum\limits_{\ell=1}^L \sum\limits_{n,n'}
\vert n\rangle\langle n\vert a\rangle\langle a
\vert n'\rangle\langle n'\vert \exp [ -i \ell (\phi_{n'}-\phi_n)]
\alpha_n^a (\alpha_{n'}^{ a})^{\ast}  \nonumber \\
& = & \sum\limits_n p_n^a \vert n \rangle\langle n\vert \,.
\end{eqnarray}

Time averaged flow from state  $\vert a \rangle$ to  $\vert b \rangle$
(or {\it vice-versa}) is governed by the
$p_{n}^a$ (for the nondegenerate case) as
\begin{equation}
P(a|b) = \lim\limits_{L\to\infty}{1\over L}  \sum\limits_{\ell=1}^L
 \vert \langle a\vert S^\ell\vert b\rangle\vert^2  = \sum\limits_{n=1}^N p_n^a
p_n^b \,.
\end{equation}

\subsection{Strong ergodicity}

A  form of ergodicity which is  too strict follows if we insist that
$P(a|b) = 1/N$ for all $a$ and $b$, in the semiclassical limit. (Note that
in this limit $N$, the dimension of the Hilbert space, goes to infinity.)
No variation is permitted in the $p_n^a$'s:   the
condition  $p_n^a = 1/N$ is necessary under such ``strict ergodicity'',
else e.g.
$P(a|a) > P(a|b)$ for $b \ne a$.  Even ``random'' eigenstates with
Gaussian random fluctuations in $\alpha_n^a $  fail strict
quantum ergodicity.
We therefore retreat to define {\it strong quantum ergodicity} to mean $p_n^a$
given by RMT: $p_n^a$'s must be 
given by a $\chi^2$
distribution of one or two degrees of freedom, depending on whether
$\alpha_n^a$  is real or complex, respectively. 

$P(a|b)$ is a quantum state-to-state measure of long-time transport.
Transport involving operators must also be considered.
The trace of an operator $F$ is defined as
\begin{equation}
{\rm Tr} F = {1\over N} \sum\limits_b \langle b \vert F\vert b \rangle \,.
\end{equation}
For   $F$ diagonal in the physically-motivated basis $|a\rangle$, 
$F=\sum_a  |a\rangle f(a)\langle a|$. A time average of $F$ results in
\begin{equation}
\bar F = \sum\limits_n \sum\limits_a  \vert n \rangle p_n^a f(a) \langle
n\vert =
\sum\limits_n f_n \vert n \rangle\langle n\vert\,,
\end{equation}
with
\begin{equation}
 f_n  \equiv \sum\limits_a p_n^a f(a) = \langle n \vert F\vert n \rangle .
\end{equation}
Strict  ergodicity is equivalent to assuming
\begin{equation}
\label{er1}
 f_n   \to {\rm Tr} (F),
\end{equation}
independent of $n$, for arbitrary $F$, as $N\to\infty$.
In other words, the time average of $F$ must be evenly distributed over
the entire available space.
(This requires $p_n^a \to 1/N$, since $f(a)$ is arbitrary.
An example is $F = \sum_{a'\in A}
\vert a'\rangle \langle a' \vert$ where $A$
is some  set of states.)
If we put some restrictions on   $F$, so that the  $f(a)$ are sufficiently
distributed over
many states $a$, then strong ergodicity is sufficient to ensure
Eq.~(\ref{er1}) holds. An example is for $a$ to be the position basis,
with $f(a)$ spread over a range of lengths of the order of $\sqrt{\hbar}$.

\subsection{Weak ergodicity}

The concept of weak ergodicity applies to further restricted operators $F$,
namely those which correspond to classical symbols.
Weak ergodicity exploits the
fact that a classically defined $F$ must have
 smooth variation of $f(a)$ with $a$, as $N\to\infty$, over
macroscopic ranges.
Given weak  quantum ergodicity, Eq.~(\ref{er1}) still holds, with the
macroscopic smoothness
restriction on $f(a)$.
The smoothness
of $f(a)$ places much less demand on the individual $p_n^a$'s.
They need only average
to uniformity over macroscopic regions. This leaves considerable latitude
for the
microscopic behavior of the $p_n^a$.  We conjecture that
the underlying classical dynamics will determine the
statistical properties of non-RMT $p_n^a$.

Weak ergodicity corresponds to SCdVZ: (almost) every
stationary state is individually
ergodic when measured over a smooth
macroscopic classical distribution,
 i.e. $ \langle n \vert F\vert n \rangle = f_n \to {\rm Tr} (F)$ 
for smooth $F$ in the
semiclassical limit.

 RMT on the other hand implies that each $p_{n}^a$ should behave as an
absolute square of an
independent Gaussian variable, giving as non-trivial results
 \begin{equation}
\label{rmt} < P(a|b)^2>  = {1 \over N^2}; \ \  < P(a|a)>  = {s \over N} \,,
\end{equation}
where $P(a|a)$ (a
time-averaged autocorrelation function for non-degenerate spectra) is the
diagonal part of the $P(a|b)$
matrix, and $s$ takes the value of 2 for generic hermitian Hamiltonians
or real Hamiltonians with complex states $|a\rangle$ (GUE statistics),
and 3 for real
symmetric Hamiltonians with real states $\vert a \rangle$
(GOE statistics, appropriate for time-reversal invariant systems).  The
measure ${\cal F}=1/(N< P(a|a)>)$ is
the fraction of the eligible eigenstates with which a typical state has
significant overlap
\cite{ergodic}.
The averages in
Equation (\ref{rmt}) are taken over all states, as well as over an
appropriate ensemble of systems.
 $< P(a|a)> $, in addition to being a long-time
averaged autocorrelation function,
also
measures the first non-trivial moment of the
$p_n^a$ distribution.
${\cal P}  = < P(a|b)^2> $ measures  state-to-state
transport fluctuations, at long times.
An
eigenstate-specific statistic can also be defined:
\begin{equation}
P(n|n) = \sum\limits_a (p_n^a)^2,
\end{equation}
which measures the inverse participation ratio for that eigenstate, that
is, the inverse of the
number of states $|a\rangle $ with which the eigenstate $|n\rangle $ has
substantial overlap. The mean of the $P(n|n)$
distribution is obviously equal to that of $P(a|a)$.
Thus, RMT implies (up
to a factor of 2 or 3 arising from
quantum fluctuations) ergodicity of eigenstates projected onto individual
quantum states, as measured by ${\cal F}$ (as well as ergodic
transport at the single-state level, as measured by ${\cal P}$,
which, as will be
shown explicitly later on,
is in fact a weaker
condition). In this paper,   deviations from single-state ergodicity are
examined which are still consistent
with the state-averaged Schnirelman results.

\section{Examples of Localization}

We will briefly discuss a few known examples of localization
phenomena, putting them in the language presented above. By ``localization"
here is meant an anomalously large
expectation value $N< P(a|a)> $ or $N^2< P(a|b)^2> $, compared with the
RMT predictions.
Examples include
scarring, weak localization, and dynamical localization.
Most of these localization phenomena
disappear or become much less important as $\hbar\to 0$. The effects we 
study here become  {\it increasingly dramatic} as  $\hbar\to 0$.

Scarring is an enhancement of the return probability $P(a|a)$ associated
with an unstable periodic orbit in
the corresponding classical system. It is a striking example of the
influence of short-time classical dynamics
on long-time quantum properties in classically chaotic systems. It can be
shown by semiclassical arguments
(and assuming randomness in the homoclinic tangle associated with the
periodic orbit), that there is a
universal scarring enhancement factor $\alpha(\lambda)$ for an initial
state lying on an unstable periodic
orbit with exponent $\lambda$. Thus, when averaging over an appropriate
ensemble,
\begin{equation} < P(a|a)>  = {\alpha(\lambda) s \over N} \,,
\end{equation} where $\alpha(\lambda)$ is a
function that decreases to $1$ as $\lambda \rightarrow \infty$, $s$ is the
RMT quantum fluctuation factor mentioned above, and
$|a\rangle $ is a coherent state centered on a periodic point. However,
significant enhancement is found only for
those initial states $|a\rangle $ which lie close to a periodic orbit with
a small amount of instability, and these
comprise a vanishing fraction of the total state space in the semiclassical
limit. Thus quantities such as
$\langle \sum_a P(a|a)\rangle $ will still be generic as $h \rightarrow 0$.

Weak localization is a (factor of $2$) backscattering enhancement in
time-reversal invariant systems which can
be understood semiclassically as resulting from the constructive
interference between any given returning path
and its time-reversed image. This enhancement factor is a constant over all
incoming states, is preserved as
$h \rightarrow 0$, and gives rise to a corresponding enhancement in the
inverse participation ratios $P(n|n)$
of the S-matrix eigenstates.

Finite enhancement factors for selected $P(a|a)$ are also found in systems
with a discrete symmetry group.
Anomalous autocorrelation statistics arise for initial states which are
eigenstates of the relevant symmetry
transformation. This effect can combine with scarring to produce more
eigenstate localization in symmetric
chaotic systems such as the stadium billiard than would be expected based
on scarring arguments alone\cite{steve}.

Finally, dynamical localization is a strong localization effect in the
sense that it can produce arbitrarily
large enhancement of the $< P(a|a)> $ (and also of the $< P(a|b)^2> $)
statistics,
for a given value of Planck's
constant $h$. So, for example, the quantum kicked rotor (standard map)
allows classical diffusion to infinite
momenta for sufficiently large values of the kicking parameter, but the
quantum eigenstates remain localized
in momentum space. The localization however does not survive the $h \to
0$ limit, in the sense
that the localization length becomes infinite.

The ``slow ergodic" systems   considered towards the end of this paper
present a new phenomenon not present in
any of the localization examples given above, in that the eigenstates
become {\it more and more localized} (in
the sense of a diverging $N< P(a|a)> $ statistic), as $h \rightarrow 0$.
However, the $N^2< P(a|b)^2> $
eigenstate-averaged transport statistic
seems to converge to a constant in this
limit. Before describing the detailed
properties of these unusual quantum systems, it is shown how such anomalies
are mathematically possible
in the formalism  outlined above.

\section{Failing strong ergodicity}

Since the Schnirelman test effectively averages eigenstates in the
semiclassical limit over a finite region in
the classical phase space (which includes infinitely many wavelengths in
this limit), it is not necessary for
eigenstate ergodicity to obtain at scales of the order of Planck's
constant. In fact, for eigenstates to be
ergodic on a classically fixed mesh as $N \rightarrow \infty$, it is
sufficient for the number of states in
which a typical eigenstate lives (the inverse of the $< P(a|a)> $
statistic) to be an increasing function of
$N$, provided the $1/< P(a|a)> $ states are spread evenly enough over the
classical phase space. There is no
requirement for $1/< P(a|a)> $ to be a linear function of $N$. Notice that
if this function is slower than
linear, the eigenstates become less and less ergodic at the individual
state level as $h \rightarrow 0$ ($N
\rightarrow \infty$), even as they appear more and more smooth on classical
scales.

What about the transport statistic $N^2< P(a|b)^2> $? Assuming the elements
$p_n^a$ are statistically
independent (but not necessarily Gaussian) variables  gives
the following relationship between $N^2< P(a|b)^2> $ and $N< P(a|a)>$:
\begin{eqnarray}
\label{constr}
N^2< P(a|b)^2>  &=& 1 + {2 \over N} \left (N< P(a|a)> \right )^2 +
{1 \over N^2} < N^4 (p_n^a)^4 >  \nonumber \\
&\equiv& C_1 + C_2 (N< P(a|a)> )^2 +
C_3  < N^4 (p_n^a)^4 > \,.
\end{eqnarray}
(Recall that $N< P(a|a)> $ measures the second moment of the $p_n^a$
distribution.)
The second term arises from the
$a=b$ and $n=n'$ contributions to the sum on the left hand side. It is apparent
that $N^2< P(a|b)^2> $ is
forced to become anomalous if $N< P(a|a)> $ grows faster than $N^{1/2}$.
On the other hand, if $N< P(a|a)>  \leq
O(N^{1/2})$, then the anomalous nature of the individual eigenstates need
not qualitatively affect the
transport properties of the system, even at the single-state level. This
is a key point
of this paper.

In reality, there will always be non-trivial correlations $< p_n^a p_{n'}^b
> $
in the eigenstate matrix for $|a\rangle $
and $|b\rangle $ connected by short classical paths, as well as generally for
$|a\rangle $ close to $|b\rangle $ (because of
diffractive effects). Such classical and quantum spreading tends to
increase $C_2$
in Eq. (\ref{constr}), imposing more stringent limits on the behavior of $N<
P(a|a)> $ which is consistent with
single-state transport remaining ergodic \footnote{It is intuitively clear
that increasing the amount of
transport in the classical system cannot make quantum transport {\it less}
ergodic -- it can, however, make
the eigenstates {\it more} ergodic, {\it i.e.} less anomalous.}. Non-trivial
classical and diffractive
transport characteristics can also change $C_1$ in Eq. (\ref{constr}) from the
RMT prediction of
$1$. This, however, is a less dramatic form of localization, because this
deviation from RMT predictions does not become ever more pronounced in the
semiclassical limit.

\section{Classical Ergodicity}

Classical ergodicity (time average equals phase space average)
implies that each
phase-space ball of radius $\epsilon$
\footnote{Or an x- or p-strip of width $\epsilon$. The rates of spreading
may well be different in different
coordinates, so that quantum ergodic properties are sensitive to the choice
of basis.} should evenly cover all of phase space under the dynamics,
thus intersecting every
other ball of the same radius after being evolved for a sufficiently long
time. One may ask how the
time $T$ required to achieve this uniform coverage
scales with $\epsilon$. This allows for a hierarchy
within the class of ergodic systems. For a purely mixing (chaotic) system, such
as the baker's map or the stadium billiard,
$T$ scales logarithmically with $\epsilon$; in other words, after a finite
time, phase space has been explored
on an exponentially fine scale. For an ergodic system with zero Lyapunov
exponent, $T$ and $\epsilon$ are
related by a power law (the power may depend on the basis under
consideration, as well as on the details of the
system). Examples of such systems include generic polygonal two-dimensional
billiards, and kicked
one-dimensional systems with piecewise-linear periodic potentials. A ``slow
ergodic" system is one for
which $T$ scales exponentially with $\epsilon$ -- for this class of systems
it takes an exponentially long
time to explore phase space evenly on a mesh of a given scale. This will
have profound consequences for the
quantum versions of these systems, in that the number of states classically
accessed by the Heisenberg time
(which is a natural cutoff time for the quantum mechanics) can be a small
fraction of the total number of
available states, even though the system is classically ergodic.

This behavior is in stark contrast with mixing systems, in which an
exponentially large number of classical
paths connect any two quantum states by the Heisenberg time. In mixing
systems, interference effects have to
work against classical spreading to produce deviations from RMT behavior
\footnote{``Hard quantum" effects
such as diffraction and tunneling generically will only {\it add} to the
rate of spreading.}. Such
interference produces the weak localization and ``discrete symmetry"
localization effects described in a
previous section, but it only leads to at most a factor of order $1$
deviation from RMT predictions. This line
of argument also suggests that diffraction may be less important in hard
chaotic systems than in non-mixing
systems, because semiclassical dynamics already produces very efficient
wavepacket spreading and transport,
helping to explain why semiclassical methods do such a surprisingly good
job of reproducing quantum mechanical
spectra and eigenstates in systems like the baker's map.

In the ``slow ergodic" systems, on the other hand, only a very small number
of states are accessed
classically by the Heisenberg time, so the non-classical effects
(diffraction, in particular), are essential
to understanding even qualitatively the quantum mechanical properties of
these systems at the single-state
level. These systems may well prove to be the best examples of non-universal
quantum ergodic behavior.

In ``ordinary ergodic" systems, characterized by a power-law spreading
rate, the situation is intermediate
between these two extremes. In such systems, there may be either of order
$1$ or a power of $N$ (depending on
the system as well as on the basis chosen) classical paths leading from a
given state to any other state
by the Heisenberg time. The effectiveness of semiclassical methods (as well
as the validity of RMT
predictions) is expected to be intermediate between the cases described above.

\section{Examples}

\subsection{Tilted Billiard}

The first example of a ``slow ergodic" system is the ``tilted billiard,"
constructed
by taking a corridor formed by two horizontal walls (in two dimensions) a
distance
$2 \pi$ apart, and closing off the right end with a
wall segment placed at some (small) angle $\alpha$ to the vertical, the
other end being left open (for now).
A classical particle, sent in from the left end at an angle $\theta$ from
the horizontal,   hits
  the tilted wall, returning at a new angle $\theta'$ (see Fig.~(\ref{tilt})).

 For angles
$\theta'$ far from the vertical, $\theta'$ is (up to a minus sign) given by
$\theta \pm 2\alpha$, depending on
whether the billiard hits the tilted wall from above or below. Near the
vertical, a double bounce off of the
titled wall can lead to $|\theta'| = \theta \pm 4\alpha$.   The angle can
always be taken
to be in the first or
fourth quadrant; then the double bounce allows the motion in angle space to
``wrap around" from $\theta =
\pi/2 - \epsilon$ to $\theta'=-\pi/2+4\alpha-\epsilon$
(for $\alpha <  \epsilon < 3\alpha$), and similarly in
the other direction.

Now   a vertical wall can be set up on the left at a distance $d$ from the
tilted wall, so that the process of
bouncing off the tilted wall is iterated (consider a surface of section
map, taking the surface to be
the left wall).
Then for irrational $\alpha/\pi$, all values of $\theta$ will be accessed
eventually for a
generic initial condition. We find, however, that the motion in
$\theta$-space is {\it not} a random walk,
which would be the case if the probabilities of the billiard ball hitting
the tilted wall from above or below
were independent of hits on previous iterations. Instead, strong
correlations exist, producing ``bottlenecks"
at certain vlaues $\theta$ for any given $d$. Near these bottlenecks, the
particle is much more likely to hit
the titled wall from below if it hit the wall from above on the previous
iteration, and vice versa.
\begin{figure}
\vskip .2in
\caption{Tilted billiard and definition of angles.}
\label{tilt}
\end{figure}
The net result is that the number of angles
visited after $T$ lengthwise traversals
of the billiard scales only as
$\log(T)$, in contrast to $\sqrt T$, which would be the consequence of a
random walk process. Similarly, the
probability of returning to a given angle after $T$ iterations, scales as
$1/\log(T)$. Specifically, if
 the typical number of angles visited (averaged over an ensemble of values
for $d$
and $\alpha$, and also over
initial conditions) is taken to be $n_T$, and the inverse of the average
probability of returning to a given angle is denoted
by $m_T$, then we find 
\begin{eqnarray}
\label{logbeh} n_T & = & r+s \log(T) \\ m_T & = & r'+s'\log(T) \nonumber \,,
\end{eqnarray}
with a numerical fit to $m_T = 1.65+0.8\log(T)$ and $n_T=1.95+0.55\log(T)$,
which works for $T$ ranging at least up to 100000 scattering events.
For the numerical calculation, we use an ensemble of billiards with 
billiard length $d$
ranging from $10\pi$ to $20\pi$, and tilt angle $\alpha$ between
$\pi/20$ and $\pi/10$. For each billiard configuartion, an average over
starting points is performed.

 This system is quantized, and  we find (numerically)
the S-matrix (at a given energy) for right-moving waves starting
at the left end of the billiard colliding with the tilted wall.  The
states which label the
S-matrix are as usual channels
characterized by the ``asymptotic'' eigenmodes in the parellel-wall
corridor, i.e.
$$\psi_n(y) = \sqrt{2\over   \pi}\sin(n   y),$$
$n=1,2,\dots$.
  For a rectangular billiard ($\alpha=0$), the S-matrix is diagonal; for
non-zero $\alpha$, a
band-diagonal structure is present, the width of the band being a fraction
of order $\alpha$ of the total
number of states (Fig.~(\ref{S})). Classically, the $n^{th}$ scattering
channel
corresponds to two incoming angles, $\pm \theta_n$, where
$$
\theta_n =  \tan^{-1}\left [ n\over\sqrt{ 8E-n^2}  \right ] \,.
$$

\begin{figure}
\vskip .2in
\caption{The S-matrix (absolute value of the $96\times 96$
martrix) for $d=2 \pi$, $\alpha = \pi/36.3777$, and $E=1120.85$ ($m=\hbar=1$).}
\label{S}
\end{figure}

In particular, if   the phases of the incoming and outgoing plane waves are
fixed, by fixing
the billiard length $d$, we can diagonalize the S-matrix and consider the
properties of the S-matrix
eigenstates. Eigenstates with eigenvalue $-1$ are in fact also eigenstates
of the closed billiard; other
S-matrix eigenstates correspond to unusual boundary conditions in the
closed system, where the wave acquires
an arbitrary phase upon bouncing off of the vertical wall.
This is actually a way of finding eigenvalues and
eigenstates of the billiard with Dirichlet boundary conditions; they are
given by eigenstates of $S$ with eigenvalue $-1$\cite{smil,bogo}.
 However here we do not seek the
  Dirichlet solutions, since they are not  special as far as their
localization
in the channel space (this has been tested numerically).  Two typical
$S$-matrix eigenstates are
shown in Fig.~(\ref{states}); these show fairly obvious non-statistical mixing
of different directions of propagation in the billiard (nonmixing of
channels in the
scattering approach).

\begin{figure}
\vskip .2in
\caption{Showing two typical $S$-matrix eigenstates for   the parameters
given in Fig.~(2).}
\label{states}
\end{figure}

\noindent Hitting the wall with positive
$\theta$ results in the next hit being at two possible angles,
namely $\pm (\theta +2 \alpha)$;  hitting the wall with negative
$\theta$ results in the next hit also being at two possible angles,
namely $\pm (\theta -2 \alpha)$.  Each channel scatters
to the vicinity of two new channels, usually one with higher angles and one
with lower angles of approach to the tilted wall. (The exceptions are
at extreme values of the angle, near 
$0$ and $\pm \pi/2$.  For example at low incident
angle (low $n$) the
new classically-accessible channels  are both higher in angle).
The $S$ and the $P(a|b)$ matrices in the channel basis  display properties
stemming from the classical behavior
of this system. In particular, $S$ (see Fig.~(\ref{S})) shows the classical
imprint
of the once iterated collision. Figure~(\ref{SS}) shows the iterations of $S$,
ending in the time-averaged
$P(a|b)$ matrix at the lower right. A detail of  $P(a|b)$ is
shown in Fig.~(\ref{pab}).

\begin{figure}
\vskip .2in
\caption{Showing the iterations of $S$ (absolute value of $S$ displayed as
density); the $P(a|b)$ matrix is shown at the lower right, for the parameters
given in Fig.~(2).}
\label{SS}
\end{figure}

\begin{figure}
\vskip .2in
\caption{Detail of the $P(a|b)$ matrix, for the parameters
given in Fig.~(2). }
\label{pab}
\end{figure}

The bands in $S$  (Fig.~(\ref{S})) are not one but two or three channels
wide, falling
off more rapidly from there.
Typically (for generic $d$, $\alpha$, and energy $E$), $P(a|b)$ has of order
$\log(N)$ strong  bands approximately parallel and in some
places perpendicular to the diagonal,
corresponding to the number of angles classically accessible,
in accordance with Eq.~(\ref{logbeh}), by the
Heisenberg time
(which scales linearly with $N$).
If all the strength were concentrated in these diagonal
lines, then both $N< P(a|a)> $ and $N^2< P(a|b)^2> $  would increase
as $N/\log(N)$ in the classical limit, giving rise to strong localization.
This semiclassical prediction is modified, however, due to the presence of
diffractive effects.
The main source of diffraction in
this system follows a roughly power-law behavior away from the classically
allowed states,
\begin{equation}
\label{diffr}
P(a|b) \sim {c_\beta \over Nn} \sum_{j=1}^n \left ({N \over b-j}\right
)^\beta\,,
\end{equation}
 where
$j$ runs over states classically accessible from $a$ by the Heisenberg
time, $n$ is the number of such
states (it scales as $\log(N)$ in the ``slow ergodic" systems under
consideration), and $\beta$ is an exponent. Because the diffractive pattern is
scale-invariant,
$N^2< P(a|b)^2> $ is
an $N$-independent constant, though   $\beta\leq 1/2$ is required for this
second moment to be normalizable.
Thus, this measure of single-state long-time transport is only off by a
constant from the RMT prediction. In fact, because of diffraction,
$N^2< P(a|b)^2> $ will not grow linearly with $N$ even in a classically
nonergodic version of this system, where the tilt angle is taken
to be rational (so that only a finite number of angles are classically
accessible).
$N< P(a|a)> $, on the other hand, exhibits a growth consistent with $\sqrt
N/\log(\nu N)$ (corresponding to
$\beta=1/2$ above). The inverse participation ratio $N< P(a|a)> $
would scale as $\sqrt N$ in the case of a rational tilt angle, where the
logarithmic classical spreading is absent.

\vskip .2in

\begin{figure}
\caption{Fit to the form given in Eq.~(17).}
\label{fit}
\end{figure}

In Fig.~(\ref{fit}),   a fit to this
form is given.
 For an ensemble consisting of 30 tilted wall billiards (using three angles
$\theta  = 0.0692, 0.0759, 0.0845$, and ten lengths 
spaced equally
from $2 \pi-4/5$
to $2 \pi +1$) the  $N < P(a|a)>$ measure is shown as a function of the
number of channels, from 9 channels to about 160.  The data fit the form
\begin{equation}
\label{fitt}
N < P(a|a)> =    {b_1\sqrt N\over b_2 + \log N}
\end{equation}
very nicely, with $b_1 = 2.87$ and $b_2 = -0.676$.
  We find that the transport $N^2 < P(a|b)^2>$ approaches a
value of about 3 in the high energy range, which is consistent with ergodic
transport.
This is an example of a system whose
existence was hypothesized earlier, with ergodic (up to a constant)
transport, but very non-ergodic individual
eigenstates, in the semiclassical limit.

Another source of diffraction causes   diffusion of the
quantum amplitude: diffraction coming from the
corners.  A sharp corner is scale invariant, so a given amount of
diffracted amplitude  emanates from the corner at each iteration as if it
were nearly a point source,
independent of $N \sim \sqrt{8E}$.
This sprays amplitude  $\sim \gamma/N$ into each channel, where $\gamma$ is
some
small number.  As $N\to\infty$, the total
corner diffracted probability scales as $\sum_{n=1}^N \gamma^2N^{-2} =
\gamma^2/N$ at each iteration.
Since there are $N$ iterations to the Heisenberg time, and assuming random
phases at each iteration,
we have $\sqrt{N} \gamma/N$ amplitude in each state due to corner
diffraction at the Heisenberg time, or a probability
averaging $\gamma^2/N$.  If this were the only source of flow between channels
the result would be localization in channel space for small $\gamma < 1$
(in practice
$\gamma << 1$); then $N < P(a|a)> \sim N$
and the transport would be anomalous,
as discussed above.  The power law channel diffraction discussed above
dominates the corner diffraction.

\subsection{Sawtooth map}

A simple example of a ``slow ergodic" system exists also within the domain
of (time-dependent) one-dimensional
models. The classical kicked sawtooth potential (KSP) map is a
discrete-time area-preserving map of the plane
onto itself, defined as follows: \begin{eqnarray} p' & = & p + a,\; x\,{\rm
mod}\,1< 1/2 \\  & & p - a,\;
x\,{\rm mod}\,1> 1/2 \\ x' & = & x + bp'\,.
\end{eqnarray} Here $a$ is the kicking strength, while $b$ measures the
free evolution time between kicks
(where the mass is set to $1$). We call this a sawtooth potential map
\footnote{Classically, this system is
very similar to a particle bouncing back and forth in a hard-wall
one-dimensional box, with periodic kicks of
fixed impulse directed towards the right side of the box.} because the
evolution in momentum can be re-written
as \begin{equation} p' = p - V'(x) \Delta t\,,
\end{equation} where $V(x)$ is a periodic sawtooth potential
\begin{eqnarray} V(x) & = & -{a\over \Delta
t}(x \,{\rm mod}\, 1),\; x\,{\rm mod}\,1< 1/2 \\  & & {a\over \Delta t}(x
\,{\rm mod}\, 1 -1),\; x\,{\rm
mod}\,1> 1/2\,, \end{eqnarray} which is
turned on for an infinitesimal time $\Delta t$, and where   the limit
$\Delta t \rightarrow 0$ is taken. We will
want to compactify this map onto a torus so as to be able to quantize it
later using a finite-dimensional
Hilbert space. We obtain an area-preserving map of the unit square onto itself
\footnote{Compactifying the $x$ variable on $[0,1]$ is natural given that
the potential is already periodic.
It then makes most sense to compactify $p$ on the interval $[0,1/b]$, in
which case   without loss of
generality   $b=1$. We may also retain $b$ as a free parameter 	while
compactifying $p$ on the interval
$[0,1]$, resulting in a discontinuous kinetic term. Fixing $b$ or allowing
it to vary has no qualitative
impact on the quantum results.}
\begin{eqnarray}
p' & =
& (p + a)\,{\rm mod}\,1,\, x< 1/2 \\  & & (p - a)\,{\rm mod}\,1,\,
x> 1/2 \\ x' & = & (x + bp') {\rm mod}\,1\,.
\end{eqnarray}

This system can be quantized in the usual way, the discrete quantum time
evolution being given by the unitary
operator \begin{equation} U = e^{-iW(\hat{p})/\hbar} e^{-iV(\hat{x})/\hbar}
\end{equation} where $W(p)$ is a kinetic term given by
\begin{equation} W(p) = -{1 \over 2} b (p\, mod \,1)^2
\end{equation}

Note that if $b=1$, then equivalently
we may take $W(p)=-{1 \over 2} b p^2$. This
would result in a different
quantization corresponding to the same classical dynamics.

When quantizing evolution on a torus, a choice of boundary conditions needs
to be made. Periodic boundary
conditions are most natural, but more generally one can select arbitrary
phases $\theta_1$ and $\theta_2$
associated with winding around the x-direction and p-direction,
respectively. The pair $(\theta_1,\theta_2)$
can also be seen as fixing (quantizing) the locations of the finitely many
($N=1/h$) p-states and x-states,
respectively, on the unit square. Physically, one can imagine varying
boundary conditions by changing the
(shielded) magnetic flux through a circle on which an electrically charged
particle is constrained to move.
One thus obtains an infinite set of quantizations, all having the same
semiclassical limit.

The kinematics on a torus is given by a basis of $N$ position states
$|i\rangle $, $i=0..N-1$, with positions
$x(i)=(i+\theta_2)/N$, and a momentum basis $|\tilde{j}\rangle $,
$p(\tilde{j})=(\tilde{j}+\theta_1)/N$ related to
this by a discrete fourier transform. We will
be interested in the properties of
eigenstates of this system in the momentum basis, which is the eigenbasis
in the $a \rightarrow 0$ limit.

First  the classical properties of this system need to be discussed. An
important property, which it shares
with the titled billiard described earlier is the logarithmically slow
spreading in momentum space.
Classically, the momentum jumps up or down by $a$ (mod 1) on every iteration,
but, as in the case of the tilted
billiard, the long-time behavior does not resemble a random walk, but
rather is logarithmic, in accordance
with Equation (\ref{logbeh}). The numerical values are
$m_T = 1.9 + 0.85 \log(T)$ and $n_T = 1.8 + 1.2 \log(T)$. In obtaining these
values,   an ensemble
average over kick strength  $a$ and kicking periodicity $b$ is performed,
as well as
an averaging over the starting point. Very
similar behavior is obtained fixing $b=1$ (in which case the kinetic term
is continuous). In this case the
numerical results are $m_T = 1.55 + 0.85 \log(T)$ and $n_T = 1.7 + 0.95
\log(T)$. 
The logarithmic behavior persists to times as long as
$T=10^7$ iterations, as shown in Fig.~(\ref{claslog}).

\begin{figure}
\vskip .2in
\caption{Logarithmic classical spreading in momentum
for kicked sawtooth map, Eq.~(15).}
\label{claslog}
\end{figure}

Diffraction has a similar effect here as in the tilted billiard system,
causing $N< P(a|a)> $ to grow only as
$\sqrt N/\log(N)$ instead of the semiclassically predicted $N/\log(N)$
behavior, while $N^2< P(a|b)^2> $ is
generic (independent of $N$).
Here averaging is performed over boundary conditions
$\theta_{1,2}$ as well as over the classical
kick strength $a$ described above (kick periodicity $b$ is fixed at $1$).
Specifically, the numerical results we
obtain are $N< P(a|a)> = 2.5\sqrt(N)/\log(2N)$ for
$N$ in the range $100$ to $1000$, while $N^2< P(a|b)^2> $ remains
constant around $1.55$ in the same range.
The power-law scaling of the diffraction pattern, in
accordance with Eq.~(\ref{diffr}), is shown in Fig.~(\ref{diffrplot}),
for a classically nonergodic version of this system, with a rational
kick strength $a=1/3$. (Using
a rational kick strength allows us to look at diffractive behavior
in isolation, in the absence of long-time classical spreading.)

\begin{figure}
\vskip .2in
\caption{Iterations of the propagator (absolute value) and the $P(a|b)$ matrix,
for kicked sawtooth map.}
\label{SSkick}
\end{figure}

\begin{figure}
\vskip .2in
\caption{Inverse participation ratio, Eq.~(17), for
kicked sawtooth map.}
\label{fitkick}
\end{figure}

This behavior is interesting in connection with the limits discussed
earlier given by Equation (\ref{constr}),
where we saw that single-state transport can remain generic as long as $N<
P(a|a)> $ grows no faster than
$\sqrt N$. In both of the systems   studied here, diffraction alone
limits $N< P(a|a)> $ to be within this bound. This may suggest that
non-classical effects are always sufficient
to give ergodic transport, no matter how slow the classical spreading may be.

\begin{figure}
\vskip .2in
\caption{Log of the power-law diffraction statistics for nonergodic
kicked sawtooth map, Eq.~(16). The peaks are fit to the form
${\rm log}[0.01/\protect\sqrt{n-j}]$,
where $j$ is the position of each peak
maximum.}
\label{diffrplot}
\end{figure}

\section{Conclusion}

The possible types of quantm ergodicity we have discussed are schematized in Fig.~(\ref{sum}).
Starting with the (impossible in general) strict quantum ergodicity, we show
RMT with its $\chi^2$ distributed $p_n^a$'s,  follwed by two types of weak 
quantum ergodicity. These two differ as to whether only $<P(a|a)>$ is anomalous
(Weak I) or whether $<P(a|b)^2>$ is also anomalous (Weak II).  All these fall
within SCdVZ ergodicity; only the last, localized case does not. The examples
presented here are both in the Weak I catagory.

We have seen that in classically ergodic systems where the classical rate
of spreading is sufficiently slow, quantum mechanics is only able to take
advantage of that classical phase space exploration which occurs before
the Heisenberg time (i.e. before the time at which individual eigenstates are
resolved). This gives rise to a previously unexplored localization
phenomenon in such systems, involving a growing deviation from random
matrix theory predictions as the classical limit $h \rightarrow 0$ is taken.
Because of the non-commutativity of the $h \rightarrow 0$ and 
$T \rightarrow \infty$ limits, we obtain a situation in which quantum
eigenstates become less and less ergodic even as the quantum mechanics
follows the (ergodic) classical mechanics for longer and longer times.
Importantly, there is nonetheless no disagreement with the prior
results of  Schnirelman\cite{schnirl},
Colin de Verdiere\cite{colin}, and Zelditch\cite{zel} concering
ergodicity of eigenfunctions.

These ``slow ergodic" systems also give rise to an interesting interplay
between classical and hard quantum (diffractive) phase space exploration.
The pattern of quantum transport can be well understood by superimposing
diffractive spreading on top of classically allowed motion. This
provides a new laboratory for examining the strengths and limitations
of semiclassical methods, and for separating out classically-based and
classically-forbidden quantum effects.

The ``tilted billiard" and ``kicked sawtooth potential" systems
discussed in this paper are part of a diverse spectrum of classical
systems and their quantum counterparts. This spectrum includes (1) integrable
systems, (2) nonergodic systems such as rational-angle polygonal
billiards, (3) ``slow ergodic" systems, (4) ergodic systems with zero
Lyapunov exponents (generic polygonal billiards), (5) mixed systems,
and (6) chaotic (i.e. mixing) systems. Different time scales and different
issues of quantum--classical correspondence arise in each of these
situations. Furthermore, transitions between many of the regimes given above
can be obtained by varying $\hbar$ for a given classical system.

In all of these cases, an understanding of eigenstate and transport
properties, as well as of spectral characteristics, is crucial for a
full appreciation of the quantum--classical relationships. In this paper,
we have outlined a formalism which should prove useful in many situations
for studying the deviations from universality of eigenstate and
transport statistics.

\section{Acknowledgements}

This research was supported by the Harvard Society of Fellows and by the
National Science Foundation under grant number CHE-9014555.

\begin{figure}
\vskip .2in
\caption{Schematic of the possible
types of quantum ergodicity, as shown through the qualitative
behavior of $p_n^a$ and $P(a|b)$.
 The first four are consistent with
SCdVZ theorem, the last is simply shown for contrast to illustrate
a nonergodic case.}
\label{sum}
\end{figure}

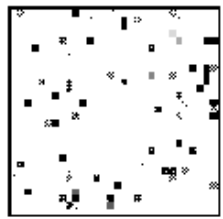
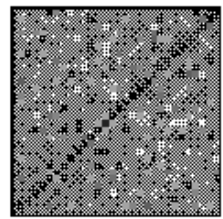
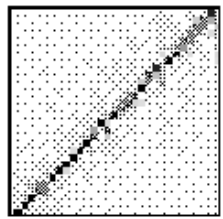
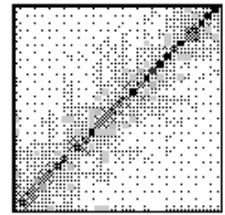
\end{document}